# AI Meets Natural Hazard Risk: A Nationwide Vulnerability Assessment of Data Centers to Natural Hazards and Power Outages

*Miguel Esparza[1*], Bo Li[1], Junwei Ma[1], Ali Mostafavi[1]*

[1] *Urban Reslience.AI Lab, Zachry Department of Civil and Environmental Engineering, Texas A&M University, College Station, TX 77840, United States of America*

**Abstract**

Our society is on the verge of a revolution powered by Artificial Intelligence (AI) technologies. With increasing advancements in AI, there is a growing expansion in data centers (DCs) serving as critical infrastructure for this new wave of technologies. This technological wave is also on a collision course with exacerbating climate hazards which raises the need for evaluating the vulnerability of DCs to various hazards. Hence, the objective of this research is to conduct a nationwide vulnerability assessment of (DCs) in the United States of America (USA). There is an increasing importance to maintain new infrastructure to support the digital age thanks to the emergence of these innovative technologies being public. DCs provide such support; however, if an unplanned disruption (like a natural hazard or power outage) occurs, the functionality of DCs are in jeopardy. Unplanned downtime in DCs cause severe economic and social repercussions. Therefore, this research uses spatial analysis methods to assess the current vulnerability of DCs toward natural hazard and power outages. With the Local Indicator of Spatial Association (LISA) test, the research found that there are a large percentage of DCs that are in non-vulnerable areas of disruption; however, there is still a notable percentage in disruption prone areas. For example, earthquakes, hurricanes, and tornadoes have the most DCs in vulnerable areas. When examining power outages, DCs reside in areas that have faced frequent power outages during 2014-2022 and experience long durations without power. After identifying these vulnerabilities, the research identified areas within the USA that have minimal vulnerabilities to both the aforementioned natural hazards and power outages with the BI-LISA test. After doing a composite vulnerability score on the Cold-Spots from the BI-LISA analysis, the research found three counties with the low vulnerability scores. These are Koochiching, Minnesota (0.091), Schoolcraft, Michigan (0.095), and Houghton, Michigan (0.096). The contribution of this research is to provide infrastructure managers with interpretable maps to guide their decision-making and by understanding the current vulnerabilities, they can develop specific solutions can to ensure the functionality of DCs.

## 1. Introduction

Our society stands on the verge of a revolution powered by artificial intelligence (AI) technologies. The exponential growth of AI has led to a significant expansion of data centers (DCs). DCs have served as critical infrastructure in the past but are becoming increasingly important to maintain due to this new technological wave powered by AI. However, this surge is on a collision course with escalating climate hazards, highlighting the urgent need to evaluate the vulnerability of DCs to various environmental threats. DCs are critical infrastructure systems that contain high quantities of servers and computers to provide digital services for many large and local businesses [1]. This research defines DCs in accordance with the U.S. Environmental Protection Agency, which is a type of principle electronic equipment that is used for processing data, storing the data and providing communication tools [2]. With the emergence of AI technology and growing dependency of societies on them, there is a dire need to examine current and future vulnerability of DCs to hazards that could cause disruptions in the functioning of these critical infrastructure. A robust DC has special power conversion and backup equipment that preserve high-quality and reliable power systems that must maintain important environmental factors [2]. In fact, the most critical systems of a DC are continuous power supply, air conditioning and internet network connectivity [3,4]. Therefore, if these systems are disrupted in an event, such as a natural hazard or a power outage, the operation of a DC is in jeopardy. Disruptions of DCs are problematic as they provide valuable communication services to business and the public [2]. During a natural hazard, communication is vital as the affected community needs to be aware of resource allocation and evacuation [5–7]. DCs also play an important role in economic activities. Even a small amount of unplanned downtime can lead to a significant loss of revenue [8]. The average cost associated with unplanned DC downtime is 8,851 USD per minute [9]. For these reasons, the functionality of a DC is critical to the built environment [10]. Recently, there has been substantial development in infrastructure systems that are driven by technological innovation and sustainable development [11]. For example, Goralski et al. (2020) highlighted AI-based technology such as Smart Water Management System and Plant Village to identify infested water which improves water infrastructure and sustainability efforts [12] . DCs are becoming increasingly more important in the modern age [13] due to the emergence of digitalization and AI [14]; therefore, robust infrastructure is needed to support these innovations. In addition, the rapid development of information and communication technologies emphasize the importance of DCs [15]. DCs provide valuable services to the emerging digital economy and promote the well-being of society; therefore, enhancing the resilience of new DC and evaluating the vulnerability of existing ones is critical. Due to the importance of DCs there is a need for a nationwide vulnerability assessment. Recognizing this gap, this study aims to assess the extent that DCs are prone to natural hazards and power outages in the United States of America (USA).

The importance of assessing risk and vulnerability of DCs has been recognized in a number of recent studies [16,17]. The growth of digital infrastructure, including DCs, is a global phenomenon. Zhang et al. (2022), used the Getis-Ord spatial statistic to compare the development of digital infrastructure and traditional infrastructure in China [11]. They found that through the China's infrastructure policy, that digital infrastructure has been growing at a rapid pace since 2018 compared to traditional systems. In addition to tracking the development of digital infrastructure, ensuring the resilience of these systems is important. Xiahou et al. (2022), applied resilience theory to the safety management of the physical components in DC infrastructure [18]. The study applies the analytic network process method in a resilience framework to evaluate the impact of natural hazards on DCs. They found that resistance capacity is the most important resilience index as the electrical systems are important to mitigate disruptions. While this framework is more quantitative, there are several qualitative methods that mitigate disruptions in DCs including linear programming [19–21]. Power outages and natural hazard impacts are some of the most severe disruptions in DC. To assess the impact of natural hazards and power outages, Ayyoub et al. (2018) used mixed integer linear programming (MILP) to model the problem of maximizing revenue for DCs and cloud service providers while considering operational constraints and disruption incidents [22]. The MILP model provides a framework for DCs managers to improve infrastructure resilience strategies, maximize revenue, and prepare for disruption events. Ju et. al (2019), assessed networks of DCs to ensure their resilience towards a natural hazard by using integer linear programming to assess different protection schemes [23]. They found that DCs that share spare capacity among different backup paths are more resilient towards a natural hazard compared to those that do not share backup paths. Identifying the placement of DCs is important to mitigate the impact of a natural hazard. Ferdousi et al. (2015) used linear programming to mitigate loss in DCs by identifying non-natural hazard-prone areas [24]. Their method accounts for the dynamic changes during a natural hazard and provides a cost analysis when employing a dynamic disaster-aware placement design. They found that accounting for natural hazards when developing data centers provides a 45% reduction of content lost and a 16.5% cost reduction. These studies emphasize the importance of considering the impact of natural hazards and power outages on DCs. However, a nation-wide study that examines the vulnerability of DCs to various hazards, as well as power outages is lacking. A nationwide study is crucial for providing a deeper understanding of the extent of susceptibility of the existing DCs and provides insights regarding regions suitable for building new DCs to support the expansion of AI technologies. The novelty of this research is using empirical data regarding natural hazards and power outages to assess the spatial extent of DC vulnerability towards these disruption events. By using spatial analysis, the research creates visual maps that are easily interpretable by infrastructure managers and the public.

The primary approach adopted in this study is spatial analysis of DCs and their intersection with spatial hazard profile of the U.S., as well as historical power outage hotspots. In the hazards and disaster literature, spatial analysis has been widely used to assess the vulnerability of critical facilities- such as grocery stores, hospitals, and pharmacies- throughout different phases of a natural hazard. For example, Dong et al. (2020) used the Local Indicator of Spatial Association (LISA) statistic to assess hospital access disruption during Hurricane Harvey in Harris County [25]. Soltani et al. (2019) used spatial analysis to assess hospital placement in Isfahan city [26]. Esmalian et al. (2022), used location-based data for a spatial network analysis to evaluate disparities in grocery store access [27]. Spatial techniques reveal patterns of infrastructure vulnerability and risk that allow decision makers to develop plans to mitigate service disruption. Hence, this study uses the LISA and BI-LISA statistic to discover patterns of DCs vulnerability towards natural hazards and power outages. This insight will allow for proper infrastructure management plans to be in place and enhance the future development of DCs.

In addition to hazards, power outages also pose significant threat to the functioning of DCs [28]. Do et al. (2023) used spatial analysis to assess power outages from 2018-2020 in the USA [29]. Through the BI-LISA analysis, they found that 62.1% of 8+ hour outages co-occurred with extreme climate related events such as heavy precipitation and tropical cyclones. Additionally, Arkansas, Louisiana, and Michigan counties experienced high outages of 8+hours in socially vulnerable areas. Other studies have examined the impact of power outages on socially vulnerable populations. Power outages not only affect these traditional infrastructure systems but DCs as well. Wu et al. (2020) formed an optimization problem that maximizes profits in DCs under power constraints. Through their simulations, they developed two algorithms that demonstrate the efficiency of maximizing profits despite power disruptions [30]. To ensure that DCs remain operational during a power outage due to a natural hazard, a Disaster Recovery Plan is assessed by Gordon et al. (2020) [31]. The researchers conducted a risk analysis and found that with proper backup cloud-based systems DCs can have a quick functionality recovery time despite disaster induced power outages. These studies demonstrate the impact that power outages have on critical infrastructure systems and the surrounding community. Therefore, this study includes power outage parameters when assessing the nation-wide vulnerability of DCs.

The increasing intensity and frequency of natural hazards along with power outages prompts the need for a nationwide DC assessment to ensure that these infrastructure systems can support the future digital society. This motivation outlines two research questions that guided this research: (1) To what extent are existing DCs vulnerable to natural hazards and power outages? and (2) What are optimal locations for DC for future development within the USA? To address the questions, the research uses the Global Moran's I, the Local Indicator of Spatial Association (LISA) and BI-LISA test to identify statistically significant clusters of census tracts in the USA that are vulnerable to natural hazard and power outages. The research will then assess the number of DC in either vulnerable (Hot-Spots) or nonvulnerable (Cold-Spots) census tracts to examine DC's susceptibility towards these events

The output of the spatial maps from research question 1 will show which census tracts in the United States have low vulnerability to natural hazards and power outages. The research will examine these areas. In addition, these area's physical traits- such as building value and agricultural value- and social

traits- such as social vulnerability and community resilience- will be examined. This will highlight areas within the USA that are suitable for future DC development.

## 2. Data and Spatial Analysis Methods

The objective of this research is to perform a nationwide vulnerability assessment on DCs towards natural hazards and power outages. The study area is the United States of America (USA). Hawaii, Alaska and the other USA territories are separated from the main analysis to maintain a proper connectivity matrix for spatial analysis. The locations of the physical DCs are from datacenter map and there are 2660 DCs in this assessment as seen in Fig. 1. Datacenter map is leading resource for DC related information as it is used by a variety of companies such as Meta, Amazon Web Service, Microsoft, and Verizon.

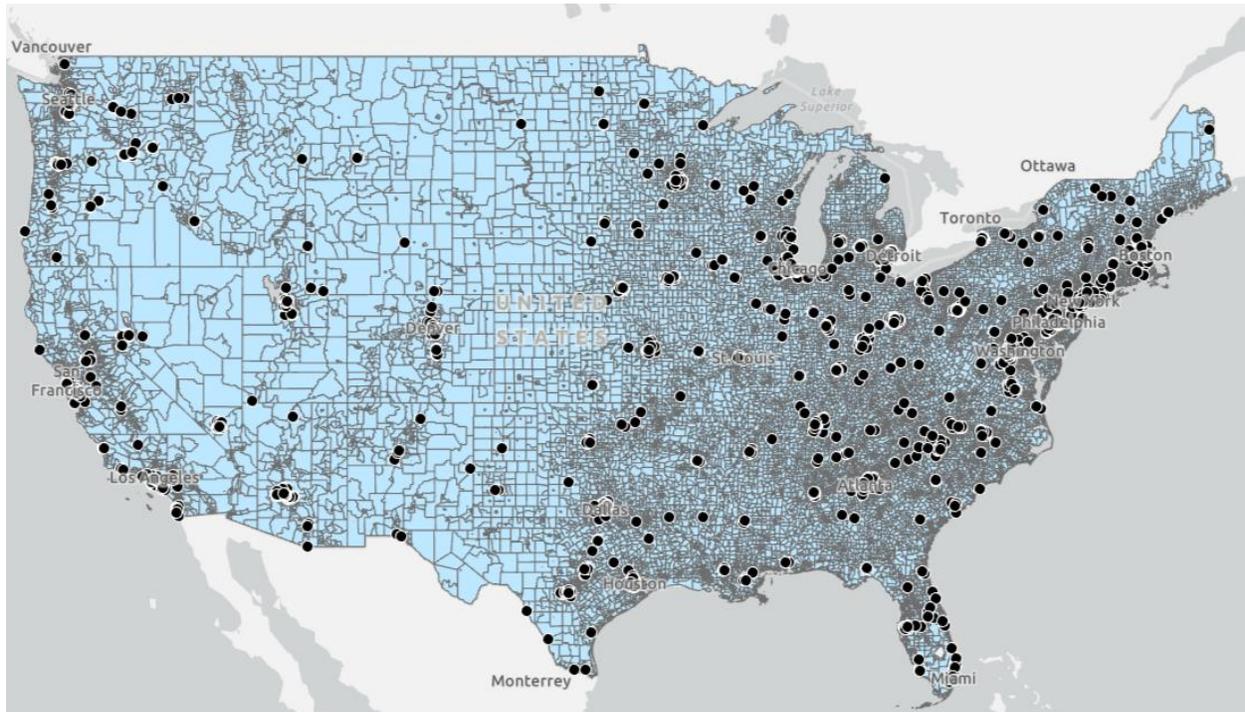

**Fig. 1. The 2660 Data Center Locations across the United States of America.**

The natural hazard data is from the Federal Emergency Management Association's (FEMA) National Risk Index (NRI). The NRI was developed by FEMA in close collaboration with various stakeholders and partners in academia; local, state, and federal government, and private industry. A series of workshops were in place to discuss proper methodology to develop the risk factor. Risk is defined by the expected annual loss due to a natural hazard while considering social vulnerability and community resilience [32]. The risk examined in this study are from earthquakes, hurricanes, tornados coastal flooding, riverine flooding, and wildfires.

To holistically assess DC vulnerability, the impact of power outages needs to be considered. This research uses power outage data from the Environment for Analysis of Geo-Located Energy Information (EAGLE-I). The EAGLE-I dataset collects power service outages, from a variety of companies, at 15- minute intervals for 3,044 out of 3,226 USA counties and county equivalents. The dataset contains the following power outage parameters: (1) the *total power outage events*, (2) the *affected customer rate*- which divides the affected customers (seconds) by total customers from the respective power outage company (3) *duration of the power outage event* (seconds), and (4) *time between outages* (seconds). The timeframe of the dataset is from 2014 to 2022. The research spatially joins the NRI and EAGLE-I dataset and uses census tracts as the spatial aggregation. The joined data will have natural hazard risk at the census tract level; however, the power outage variables are only available at the county level. This leads to multiple census tracts that reside in the same county having similar power outage parameters. This limitation is acknowledged by the research; however, a smaller spatial scale of the EAGLE-I dataset does not cover the whole USA. Another limitation with the data is that some census tracts will have missing power outage parameters. These missing values are approximated by averaging the respective power outage parameter with neighboring census tracts. The rational for this approach is to ensure that the connectivity matrix needed for the LISA and BI-LISA test is complete and does not have missing values. Despite these limitations, the need for a nation-wide vulnerability assessment of DCs is paramount to support the new technological advancements driven by AI and to ensure the longevity of innovation.

Spatial analysis can be performed once the joined dataset has NRI, EAGLE-I, and number of DCs in each census tract for USA. The spatial analysis techniques will reveal areas that are vulnerable to natural hazards or power outages (Hot-Spot) and areas that are less prone to these disruptions (Cold-Spot). When examining risk, it is important to recognize that Cold-Spots may still face disruptions due to the unpredictable nature of natural hazards; however, these areas are least likely to experience disruption from these events. The research will examine the number of DCs in either Hot-Spots or Cold-Spots, generated from the spatial analysis techniques described in section 2.1, to address research question 1. Research question 2 is addressed by comparing LISA maps to find the optimal areas that are in Cold-Spots as well as using the BI-LISA test to examine areas that are less prone to a combined affect from natural hazards and power outages. Cold-Spots from the BI-LISA test are examined further by using a composite vulnerability score as described in section 2.2. Counties with a low vulnerability score are suggested for DC development, thus providing characteristics and counties that answer research question two.

*2.1 Spatial Analysis to Assess Vulnerability Patterns in the United States of America*

Spatial analysis is performed on the combined dataset to assess the number of DCs that are in Hot-Spots and Cold-Spots. The first step of spatial analysis is to perform the Global Moran's I test to determine if neighboring census tracts share a similar experience with natural hazard and power outage vulnerability. The null hypothesis for the Global Moran's I tests holds that the attributes being analyzed are randomly distributed among the features in the study area. Therefore, a p-value of less than 0.05 indicates the emerging spatial patters are not random [33].

In the next step, the study area is decomposed into local clusters due to the statistical significance of the Global Moran's I test. To break down the global space, the LISA statistic is used for identifying clusters which have similar values of the respective disruption. Obtaining different spatial clusters through spatial analysis can help identify the extent of DC vulnerability towards natural hazards and power outages [34].

The Local Moran's I statistic allows for the decomposition of the Global Moran's I index into individual observations to assess the significance and contribution of the local clusters. The results of this test produced statistically significant polygons that form clusters of either natural hazard risk or the power outage parameters. The cluster categories are as follows:

- High–high (Hot-Spots): Areas that have a high means and are surrounded by other geographical areas with similar values.

- Low–low (Cold-Spots): Areas that have a low means and are surrounded by other geographical areas with similar values

- High–low: Areas that have a high means and are surrounded by other geographical areas with lower values

- Low–high: Areas that have a low means and are surrounded by other geographical areas with higher values

Another component of spatial analysis is using a multivariate analysis to assess the impact of both natural hazard risk and power outage vulnerability through the BI-LISA test. The BI-LISA test measures the degree which the value for a given variable at a location is correlated with its neighbors for a different value. This will allow the research to identify the best Cold-Spots for development, as well as consider the two outlier variables for DC vulnerability. The BI-LISA follows similar logic and methodology. This allows the research to examine the number of DCs that are in areas that are prone to both disaster and power outages.

*2.2 Composite Vulnerability Score to Identify areas for Data Center Development*

To address research question 2, the Cold-Spots generated from the BI-LISA analysis will be examined further. Specifically, the following features will be used in a composite score to determine vulnerability toward disruption events: Building Value, Agriculture Value, Social Vulnerability, Community Resilience, all natural hazard features, and all power outage features. The rational for including the first four features (which are from the FEMA NRI dataset) is that the physical and social components of natural hazard impact is considered in the analysis. Moreover, by including the economic impact of repairing buildings and land the research can better identify areas suitable for DC development. Additionally, including the social components allows the research to consider areas that can recover quickly from a disaster. While just Cold-Spots are being examined, all natural hazard and power outage features are considered for this analysis due to the objective of the research and this will ensure no natural hazard is overlooked. A composite score is used to rank counties due to the simplicity and interpretability of the analysis as seen in equation 1 [35]. Where, $\omega_i$ is the feature weight multiplied by the normalization of the respective feature.

$$Composite\ Vulnerability\ Score = \sum_{i=1}^{n} \omega_i \frac{x_i - \min(x_i)}{\max(x_i) - \min(x_i)} \tag{1}$$

While methods such as PCA [36–38] has been used to score spatial areas, the research determined that a normalization and composite scoring method retains the original meaning of the features and allowed the research to assign proper weights. For example, a higher weight is assigned to Earthquake Risk if the Cold-Spots being examined were generated from the Earthquake Risk Score during the BI-LISA analysis. Similar logic is applied to the power outage features. The physical and social features have less weight compared to the natural hazard and power outage features. Finally, the total sum of the weights are equal to one to ensure the ranking of the counties have a score between zero to one.

## 3. Results and Discussion

The Global Moran's I is computed for the examined natural hazard along with the power outage parameters. Once the statistical significance of the global space is established, the LISA clusters are examined to identify Hot-Spots and Cold-Spots of DC vulnerabilities. Based on the LISA results, parameters are selected for the BI-LISA test and the number of DC are assessed. Examining the number of DCs in Hot-Spot and Cold-Spots with the LISA and BI-LISA test will address research question 1. This can provide valuable information to DC owners, operators, and managers as they can be more aware of how natural hazards and power outage parameters affect this infrastructure system.

To address research question 2, the LISA cluster maps are compared among the observed natural hazards and power outage parameters. This will identify which areas have overlapping Cold-Spots that could be suitable for development. To validate these patterns, the BI-LISA test is preformed between the disruption events with the highest number of DCs in Hot-Spots from the LISA test. The rationale is that these maps will provide the most insight on DC vulnerabilities as the BI-LISA test will examine areas that are in Cold-Spots for both natural hazards and power outages. To identify the most suitable counties a composite vulnerability score is used within the Cold-Spots of the BI-LISA analysis. This will rank the least vulnerable counties towards natural hazards and power outages, while accounting the physical and social traits of the county. The results in table x will properly address research question 2, and Figure x and x of the spatial results provides a broad overview to further answer research question 2.

### 3.1 Identifying vulnerable Data Centers

To investigate DC vulnerability toward natural hazards and power outages, the spatial reach of these disruptions are examined. Table 1 shows the results of the Global Moran's I test and the number of DCs in each cluster category generated from the LISA test. The Global Moran's I is statistically significant for all events which suggest that neighboring census tracts share a similar experience with natural hazard risk. Now, the global space can be decomposed into local clusters with the LISA test to identify the number of DC that are in Hot-Spots and Cold-Spots.

**Table 1: Number of Data Centers in Spatial Clusters of Natural Hazards**

|  | Coastal Flooding | Riverine Flooding | Hurricane | Tornado | Earthquake | Wildfire |
|---|---|---|---|---|---|---|
| **Global Moran's I** | 0.792*** | 0.599** | 0.982*** | 0.908*** | 0.894*** | 0.527** |
| **LISA Results** | | | | | | |
| **High-High** | 251 | 259 | 913 | 793 | 765 | 275 |
| **Low-Low** | 1198 | 652 | 1157 | 713 | 439 | 1187 |
| **Low-High** | 37 | 15 | 0 | 0 | 0 | 10 |
| **High-Low** | 14 | 59 | 1 | 17 | 184 | 128 |
| **Not Significant** | 1169 | 1675 | 589 | 1137 | 1272 | 1060 |

*p<0.05, **p<0.01, *** p<0.001*

The natural hazard with the most DCs located in Hot-Spots are hurricanes (913, 34% of DCs), tornados (796, 29.8% of DCs), and earthquakes (782, 28.7% of DCs). Conversely the natural hazard with the most DCs located in Cold-Spots are coastal flooding (1198, 45% of DCs), hurricane (1157, 43.4% of DCs), and wildfire (1187, 44.6% of DCs). These results show that while a portion of DCs are in Cold-Spots, the threat of a natural hazard is still present. This result provides future infrastructure developers valuable information on existing vulnerabilities of DCs. Generally, DCs are developed in Cold-Spots of natural hazards. This makes intuitive sense as infrastructure managers have guidelines that provide robust information on how to develop a DC [4]. However, the threat of a natural hazard does exist and must be accounted for. To further assess DC vulnerabilities, the spatial reach of different power outage parameters are examined.

Table 2 shows that the Global Moran's I is statistically significant for all power outage parameters, therefore, the research examines the number of DCs in each cluster category generated from the LISA test. The power outage parameter with the most DCs located in Hot-Spots is the number of total power outages (502, 18% of DCs). The duration of an event has 320 (12.03% DCs) DCs located in Hot-Spots. These two parameters highlight that DCs are vulnerable to power outages and if an outage does occur DC could be at risk of prolonged down time. However, number of total outages (448, 16.8% of DC), affected customer rate (1048, 39.4% of DCs), duration of the power outage event (824, 30.9% of DC) and time between the event (816, 30.7% of DCs) mostly have DCs that reside in Cold-Spots. These results show that the total power outage event feature has the most DCs located in Hot-Spots; however, DCs are generally in Cold-Spots with low power outages. It is important to note that the limitation of this analysis is the fact that no smaller spatial scale of this data is available. Moreover, to further assess DC vulnerability the combined effect of natural hazard risk and power outage risk must be considered.

**Table 2: Number of Data Centers in Spatial Clusters of Power Outage Parameters**

|  | Total Power Outage Events | Affected Customer Rate | Duration of the Event | Time Between Outages |
|---|---|---|---|---|
| **Global Moran's I** | 0.973*** | 0.482* | 0.731** | 0.909*** |
| | | **LISA Results** | | |
| **High-High** | 502 | 82 | 320 | 102 |
| **Low-Low** | 448 | 1048 | 824 | 816 |
| **Low-High** | 3 | 3 | 1 | 2 |
| **High-Low** | 0 | 0 | 0 | 0 |
| **Not Significant** | 1707 | 1527 | 1515 | 1740 |

*$p<0.05$, **$p<0.01$, *** $p<0.001$*

To further examine the spatial reach of natural hazards and power outages, the BI-LISA test is examined. The rationale is that by looking at a combined effect we can capture areas that may have DCs in Cold-Spots for a natural hazard but Hot-Spots for power outage, or vice versa.

To keep the insights from the analysis interpretable, the research selected three natural hazards and one power outage parameter with the most Hot-Spots. The selected features are hurricane, earthquake, tornado, and total number of power outage events. This will form the following features: (1) hurricane-total power outage risk (HPO), (2) earthquake- total power outage risk (EPO), and (3) tornado-total power outage risk (TPO).

Table 3 shows the Global Moran's I and number of DC located in each cluster category for the aforementioned types of hazards and total power outage. The reason for a negative Global Moran's I for TPO due to the high number of High-Low and Low-High clusters. The results show that 765 (28.7%) DCs are located in Hot-Spots of EPO. HPO and TPO have less DC in Hot-Spots as there are 265 (0.099%) and 263 (0.098%) respectively. For Cold-Spots EPO has the most with 439 (16.5%) DCs in these areas. HPO and TPO have 395 (14.8%) and 324 DC (12.2%) in Cold-Spots respectively.

The rationale behind assessing the BI-LISA test is twofold. One, the research identifies the clusters that are less vulnerable to both natural hazards and power outage. In addition, identifying the outliers of Low-High and High-Low clusters are important as they can reveal risk to one event despite having low risk to the other event. The number of DCs that have low risk towards a natural hazard but high risk for power outage are 0 (0%) for EPO, 197 (0.074%) for HPO, and 209 (0.079%) for TPO. Moreover, the number of DCs that have high risk for a natural hazard and low risk for power outage is 184 (0.069%) for EPO, 371 (13.9%) for HPO and 441 (16.6%) for TPO.

**Table 3: Number of Data Centers in Spatial Clusters of Combined Disruption Events**

|  | Earthquake and Total Power Outage Events (EPO) | Hurricane and Total Power Outage Events (HPO) | Tornado and Total Power Outage Events (TPO) |
|---|---|---|---|
| **Global Moran's I** | 0.103** | 0.0839** | -0.187* |
| **BI-LISA Results** | | | |
| **High-High** | 765 | 265 | 263 |
| **Low-Low** | 439 | 395 | 324 |
| **Low-High** | 0 | 197 | 209 |
| **High-Low** | 184 | 371 | 441 |
| **Not Significant** | 1272 | 1432 | 1272 |

*$p<0.05$, **$p<0.01$, *** $p<0.001$

These results show that the coupled impact of both a natural hazard and power outage is present and must be considered as more risk is captured by examining both events simultaneously. In addition, EPO has the most DC located in Hot-Spots which highlights the impact that earthquakes have on DC. This is well known among infrastructure managers as seismic design of a DC is a point of emphasis; however, they must be aware of other impacts such as hurricanes or tornados in conjunction with power outage disruptions.

Table 1 through Table 3 address research question 1 as they identified earthquakes, hurricanes and tornados as the natural hazards with the most Hot-Spots. Also, most DCs are in Hot-Spots for total number of power outages and duration of the power outage event. Moreover, there is a combined effect of vulnerability with respect to a natural hazard and power outage as indicated by results in Table 3. The Low-High and High-Low clusters in Table 3 show that when examining events together there exist combined vulnerabilities.

The contribution of research question 1 to infrastructure management is that the extent of DC vulnerability is discovered and future disruptions must be mitigated for DC functionality. While DCs generally resides in Cold-Spots, there are notable DCs in Hot-Spots of disruption events. To gain further insight, section 3.2 examines the locations of the Hot-Spots and Cold-Spots to identify locations suitable for future DC development which will address research question 2.

*Section 3.2. Identifying Optimal Locations for Data Center Development*

To properly address part of research question 2, clusters of Hot-Spots and Cold-Spots are examined with the LISA test. Fig. 2 shows the LISA clusters for natural hazards. As expected, the east coast is prone to Hurricanes while the west coast is prone to earthquakes. In the middle parts of the USA there are an abundance of Hot-Spots of tornadoes and riverine flooding. However, there are Cold-Spots of hurricanes and earthquakes these areas. This shows that no one location will be safe from all types of natural hazards. Despite this, identifying the locations that are least prone to multiple hazards is important. By examining these maps, infrastructure managers can enhance the resilience of DC towards the natural hazard by knowing the type of vulnerabilities that exist in the developing areas.

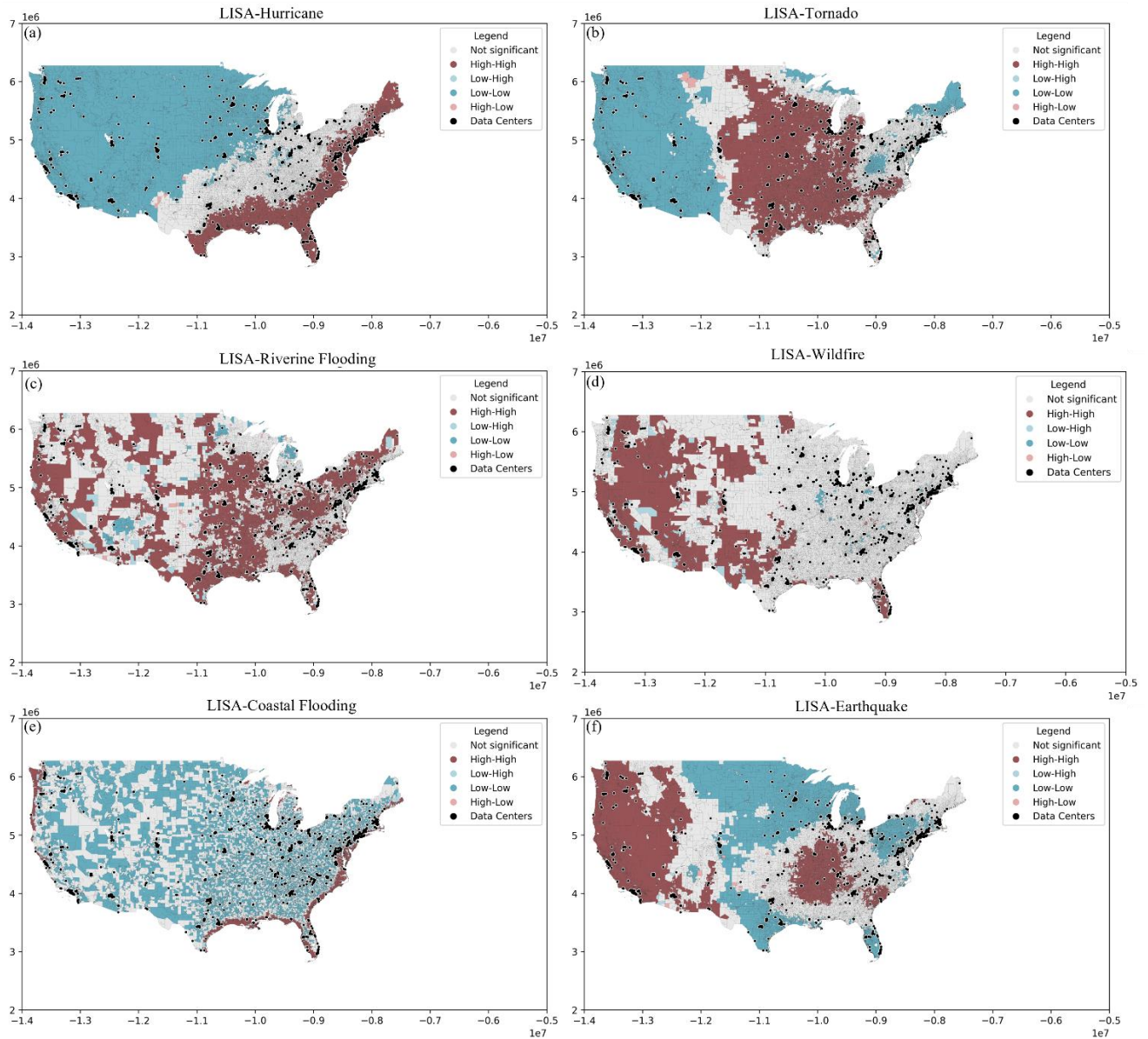

**Fig. 2. LISA Clusters of (a) Hurricanes, (b) Tornados, (c) Ravine Flooding (d) Wildfire (e) Coastal Flooding and (f) Earthquakes along with the distribution of Data Centers**

To find the optimal locations for DC development, power outage vulnerability must be considered as seen in Fig. 3. Fig. 3 shows the different power outage features. In the middle part of the USA, there are Cold-Spots of power outage events in conjunction with Hot-Spots of time between events. This shows that the low number of power outage events can be attributed to the long time between events. However, when an event does occur these areas have a prolonged duration as a cluster of Hot-Spots overlap these areas. A prolonged duration can lead to many customers being affected as indicated in Fig. 3d. This general pattern can be seen a little bit on the upper west coast. On the east coast there is a pocket of census tracts that have low time between events, prolonged duration of power outages and many affected customers. In the south of the USA there are small Hot-Spots for number of power outage events and affected customers. These are also attributed to the low time between the events.

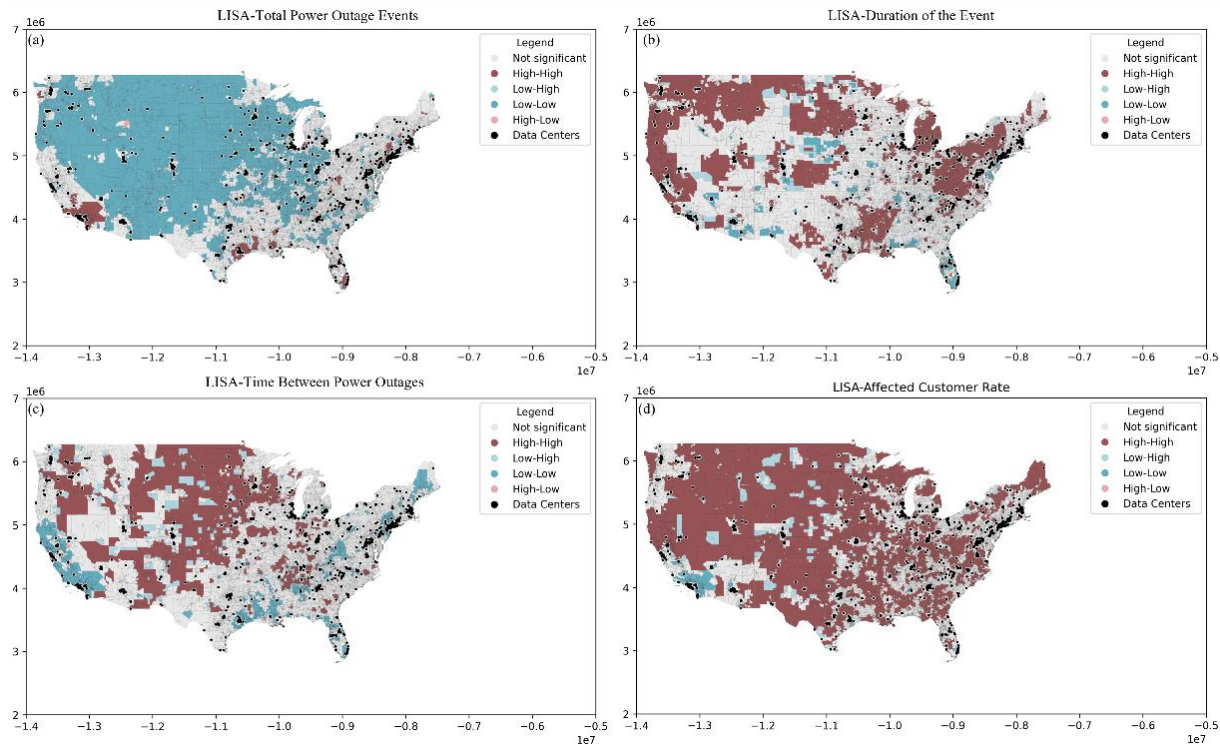

**Fig. 3.** LISA Clusters of (a) Total Power Outage Events, (b) Duration of the Outage Events, (c) Affected Customer Rate (d) Time Between Power Outage Event along with the distribution of Data Centers

By comparing Fig. 2 and Fig. 3 part of research question 2 is addressed. The middle part of the USA could be a location of interest due to the Cold-Spots of total power outage, hurricane and earthquakes. However, DC developers should be aware of the potential for riverine flooding and tornados. Moreover, if a power outage does occur there is an abundance of customers affected. Some areas that could be potential dangerous for DC development would be the in the southern/mid-south parts of the USA as there are Hot-Spots for hurricane, tornadoes, earthquakes, ravine flooding, and to an extent coastal flooding. In addition, in these areas, there are a few pockets of Hot-Spots for total power outage. In this area, there are also prolonged power outage durations when an event occurs. The LISA test, we examine overlapping patterns of hazards to provide infrastructure managers with the knowledge needed when properly developing DC. Natural hazards and power outage events are inevitable but minimizing exposure can be achieved through examining spatial maps. To further address research question 2, the BI-LISA maps are developed.

Earthquake, hurricane, tornado and number of power outage events are selected due to the number of DC in Hot-Spots from the LISA analysis. Fig. 4 shows the four different BI-LISA cluster categories outlined in section 2. Similar logic can be seen, as the middle areas have low EPO and HPO. However, there are clusters of high tornado risk but low power risk. Now that the research has identified potential locations for DC development, the composite vulnerability score is computed for the BI-LISA Cold-Spots. This will identify the most suitable counties for DC development.

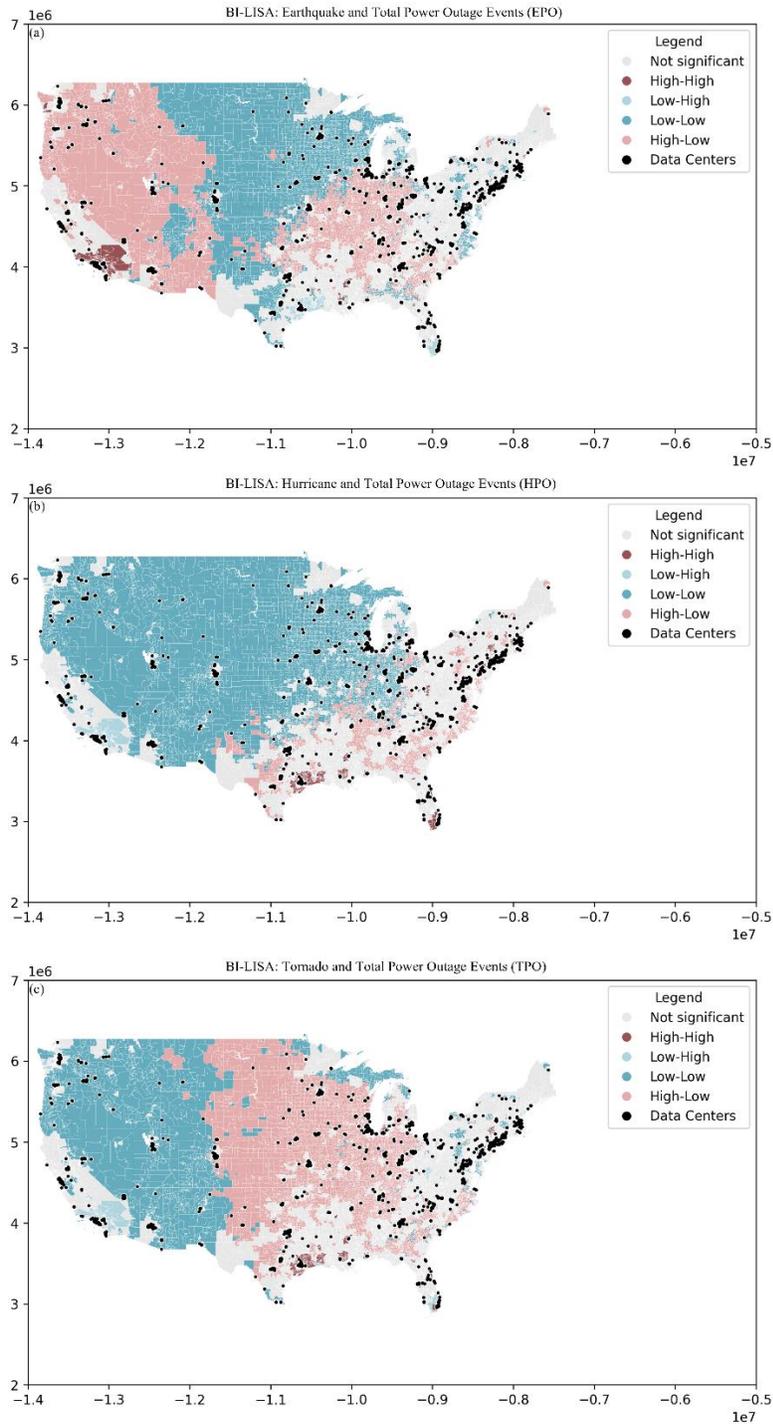

**Fig. 4.** BI-LISA Clusters of (a) Earthquake and Total Power Outage Events (EPO), (b) Hurricane and Total Power Outage Events, (c) Tornado and Total Power Outage Events (TPO) along with the distribution of Data Centers

Based on the composite vulnerability score, five counties from each Cold-Spot category are ranked in Table 4. Table 4 also shows the number of natural hazards, power outages, and DC between 2014-2022. Notable states that have multiple counties with a relatively low score are Michigan (with Schoolcraft,

Houghton, Marquette, and Luce County), and Minnesota (with St. Louis, Cook, Koochiching, and Lake of the Woods County). The counties with the top three lowest scores are Koochiching (0.091) Schoolcraft (0.095), and Houghton (0.096). Results from this table address research question 2 as these areas are in Cold-Spots for natural hazard and power outages, have little frequency of a natural hazard, and have a low composite score that considers the physical and social features. Additional information on the physical and social features are in Appendix A.

**Table 4: Composite Vulnerability Score Results**

| County, State | Score | Earthquake Events | Hurricane Events | Tornado Events | Wild Fire Events | Coastal Flooding Events | Riverine Flooding events | Total Power outages | Number of Data Centers | Cold Spot-Cluster |
|---|---|---|---|---|---|---|---|---|---|---|
| Schoolcraft, Michigan | 0.095 | 0 | 0 | 1 | 0 | 0 | 0 | 538 | 0 | HPO |
| Houghton, Michigan | 0.096 | 0 | 0 | 1 | 0 | 0 | 0 | 733 | 0 | HPO |
| St. Louis, Minnesota | 0.101 | 0 | 0 | 2 | 0 | 0 | 0 | 1585 | 0 | HPO |
| McPherson, Nebraska | 0.104 | 0 | 0 | 10 | 0 | 0 | 0 | 4 | 0 | HPO |
| Cook, Minnesota | 0.105 | 0 | 0 | 0 | 0 | 0 | 0 | 538 | 0 | HPO |
| Koochiching, Minnesota | 0.091 | 0 | 0 | 1 | 0 | 0 | 0 | 538 | 0 | EPO |
| Lake of the Woods, Minnesota | 0.103 | 0 | 0 | 6 | 0 | 0 | 0 | 565 | 0 | EPO |
| Petroleum, Montana | 0.114 | 0 | 0 | 7 | 0 | 0 | 0 | 130 | 0 | EPO |
| Marquette, Michigan | 0.115 | 0 | 0 | 1 | 0 | 0 | 0 | 614 | 0 | EPO |
| Luce, Michigan | 0.116 | 0 | 0 | 1 | 0 | 0 | 0 | 332 | 0 | EPO |
| Daggett, Utah | 0.134 | 0 | 0 | 2 | 0 | 0 | 0 | 399 | 0 | TPO |
| McKinley, New Mexico | 0.153 | 0 | 0 | 1 | 0 | 0 | 0 | 61 | 0 | TPO |
| Routt, Colorado | 0.161 | 0 | 0 | 1 | 0 | 0 | 0 | 520 | 0 | TPO |
| San Miguel, Colorado | 0.162 | 0 | 0 | 0 | 0 | 0 | 0 | 109 | 0 | TPO |
| Lewis, Idaho | 0.166 | 0 | 0 | 1 | 0 | 0 | 0 | 1273 | 0 | TPO |

Table 4 and Fig. 4 address research question 2 as they identified optimal counties that could be suitable for DC development as these areas have a low risk towards natural hazards and power outages. This will help AI technology develop and maintain operation despite disruption. Moreover, since these DC are less vulnerable towards disruption, DC developers can properly enhance the energy efficiency of these areas to keep up with the computational cost it takes to train large AI models. Ensuring the resilience of DC and maintaining their energy efficiency are crucial steps for the ethical and sustainable use of AI technology. This research contributes to the existing knowledge of DC development by utilizing empirical data from FEMA NRI and EAGLE-I in a nation-wide spatial analysis to identify the existing risk of DC and optimal locations for future development. In doing so, this research can be utilized by infrastructure managers to ensure that DC can support the increased use in AI technology in a sustainable way.

## 4. Conclusion

As we stand on the cusp of a transformative era, AI technologies are poised to revolutionize virtually every aspect of our society, especially from an economic, communication and technologic perspective. Therefore, there is a rising need to properly maintain and ensure DCs are functional after disruption events. Moreover, the rapid and exponential advancement of AI has led to an unprecedented expansion of DCs around the globe. These facilities have become the backbone of this technological surge, serving as critical infrastructure that processes, stores, and manages the vast amounts of data required for AI applications. However, this explosive growth is occurring alongside escalating climate hazards—including extreme weather events, rising sea levels, and increasing temperatures—that are intensified by climate change. This convergence sets the stage for potential conflicts, as DCs are often vulnerable to environmental threats such as flooding, hurricanes, and earthquakes. Therefore, it is imperative to thoroughly evaluate and enhance the resilience of data centers against these climate-induced risks to ensure the continuity and reliability of AI-driven services that our modern society increasingly depends upon. For these reasons, a vulnerability assessment is needed to set the groundwork for improving DC placement and enhance the resilience of these critical facilities toward disruption events.

To accomplish this need, the research uses the FEMA NRI data in conjunction with the EAGLE-I dataset to develop Hot-Spots and Cold-Spots of vulnerability at the census tract level. This study found that while a majority of DC reside in Cold-Spots, the presence of vulnerability towards natural hazards exists as an abundant of DC are in Hot-Spots of earthquake (765), hurricane (913), and tornado (793). Moreover, the total number of power outages between 2014-2022 had the most Hot-Spots (502) of DC vulnerability compared to the other power outage features this study explores. Once the spatial reach of vulnerabilities is determined, the parameters with the most Hot-Spots are used to develop the BI-LISA results. This will assess the combined effect of vulnerability and provide the research insight on locations where disruption events are low for both events. The BI-LISA analysis showed that EPO, HPO and TPO had 765, 265, and 263 DCs in Hot-Spots respectively. By examining the number of DC in Hot-Spots the study addressed research question 1 and found that generally DC are developed in Cold-Spots; however, there is still a noticeable amount of DCs in Hot-Spots of vulnerability and infrastructure managers must be mindful of this during future DC development.

To address the second research question, the spatial maps of the LISA and BI-LISA analysis are compared. Generally, the middle and northern parts of the USA could be suitable for development indicated by the Cold-Spots of EPO and HPO; however, these areas have vulnerability to tornados but low vulnerability to total number of power outages. To further assess these locations, a composite vulnerability score is calculated based on natural hazard risk, power outage vulnerability, physical features, and social features. This calculation is performed on the Cold-Spots from the BI-LISA analysis and found 15 counties that have low risk and no DC. The states with the most counties were Michigan and Minnesota. Moreover, top three counties with the lowest scores were Koochiching (Minnesota), St. Louis (Minnesota), Schoolcraft (Michigan). By examining the LISA and BI-LISA Cold-Spots, the research found suitable areas for future development.

Other studies can build off these findings by integrating more features that impact DC development such as location near other DC to enhance redundancy, energy these DC will use for sustainability in DC. In addition, future research can assess DC development on a smaller spatial scale, such as picking a study area at the county level and determine the precise location infrastructure managers can explore. The spatial scale in this study is a limitation. The EAGLE-I dataset has valuable power outage information but is at the county level. Therefore, the research had to approximate the power outage data to have the spatial scale at the census tract level. Moreover, the score calculation was simplified due to the fact that the research preformed a nation-wide analysis and the computation time of a more detailed analysis was not feasible; however, it is recommended that a more detailed score can be computed to find more insights on DC development.

Despite these limitations, a practical contribution of this study is the spatial maps generated by the LISA and BI-LISA test provide easily interpretable visuals for infrastructure managers to create proper DC development plans. They can realize what they of existing vulnerability is present in DCs. For example, if a DC is built in a tornado prone region, they can ensure the structural integrity of the DC can withstand the natural hazard. Moreover, if DCs are developed in areas with high power outages, they can ensure there are robust backup generators to ensure functionality. By having proper spatial maps, tailor made solutions can be created to ensure functionality of DCs despite disruption events. These spatial maps also allow for practical implementation of DCs as infrastructure managers can identify the existing vulnerabilities. Moreover, these spatial maps can be enhanced at a local level with smaller data aggregation to examine specific states. This will identify the precise locations that will lead to optimal DC placement. Overall, this study provides the groundwork for future vulnerability assessments that can be done at smaller spatial scales. By using empirical data, the research provides valuable information on DC development and their existing vulnerabilities. Using a spatial analysis in this study has allowed the research to determine locations at the county level that minimize DC vulnerability towards natural hazards and power outages. Proper development will allow these critical facilities to support the increase use in AI technology and provide valuable services to businesses as well as the general public.

## ACKNOWLEDGMENTS

We would like to thank Data Center Map for providing the locations of the Data Centers. Any opinions, findings, conclusions, or recommendations expressed in this research are those of the authors and do not necessarily reflect the view of the funding agencies.

## DATA AVAILABILITY

The data that support the findings of this study are available pre request from Data Center Map. The EAGLE-I data is available per request from [The Environment for Analysis of Geo-Located Energy Information's Recorded Electricity Outages 2014-2023](#).

## REFENCES

# APPENDIX

Table 1A shows the social and physical traits from NRI across the whole United States. Given these descriptive statistics, the research will place an emphasis on areas that have low building value and moderate agricultural value when examining the quantile maps, social vulnerability and community resilience will be examined as a compliment to the results from the composite vulnerability score.

**Table 4: Descriptive Statistics of Physical and Social Parameters to Assess Data Center Location**

| Descriptive Statistics | Mean | Standard Deviation | Minimum | Maximum |
|---|---|---|---|---|
| Social Vulnerability-USA | 50.13 | 28.82 | 0 | 100 |
| Social Vulnerability-EPO | 54.49 | 26.802032 | 0 | 100 |
| Social Vulnerability-HPO | 51.97 | 26.281787 | 0 | 100 |
| Social Vulnerability-TPO | 55.49 | 27.101408 | 0 | 100 |
| Community Resilience- USA | 50.16 | 28.78 | 0 | 100 |
| Community Resilience- EPO | 47.61 | 32.869533 | 0 | 100 |
| Community Resilience- HPO | 50.16 | 28.78 | 0 | 100 |
| Community Resilience- TPO | 41.624964 | 29.535686 | 0 | 100 |
| Building Value- USA | $735,621,300 | $548447000.0 | 0 | $47,179,160,000.0 |
| Building Value- EPO | $818,225,700 | $1,355,222.0 | 0 | $47,179,160,000.0 |
| Building Value- HPO | $831,174,700 | $548447000.0 | 0 | $47,179,160,000.0 |
| Building Value- TPO | $700,697,300 | $621,674,700 | 0 | $47,179,160,000.0 |
| Agriculture Value- USA | $5,299,324 | $28286050.0 | 0 | $1,875,880,000.0 |
| Agriculture Value- EPO | $25,256,180 | $86,647,470 | 0 | $1,875,880,000.0 |
| Agriculture Value- HPO | $23,037,180 | $72,896,070 | 0 | $1,875,880,000.0 |
| Agriculture Value- TPO | $9,556,422 | $42,926,760 | 0 | $1,875,880,000.0 |

Fig. 1A shows quantile maps of the physical features for the respective Cold-Spot map generated by the BI-LISA analysis and Fig. 2A shows the quantile maps of social features. Based on the descriptive statistics the research aims to assess areas with low building value that could be suitable for DC development as that parameter indicates the potential for more damage if a natural hazard occurs. When examining these figures, the maps align with the composite vulnerability score as the northern parts of the figures have desirable physical and social traits that are suitable for DC development.

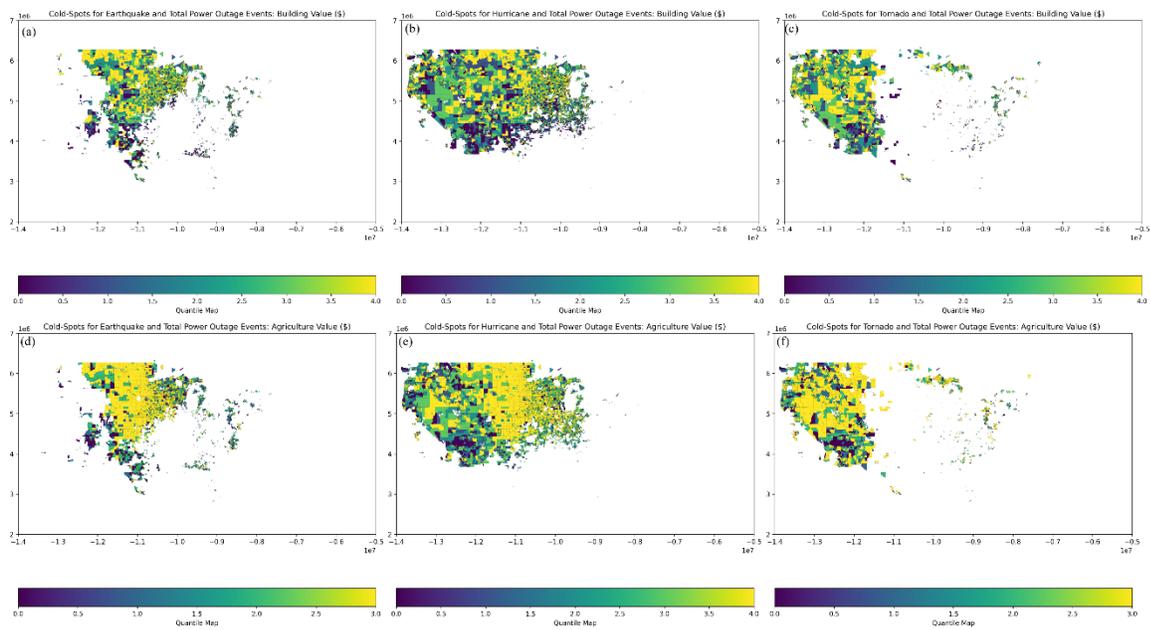

**Fig. 1A.** Building Value in Cold-Spots with respect to (a) Earthquake and Total Power Outage Events, (b) Hurricane and Total Power Outage Events, and (c) Tornado and Power Outage Events. Agricultural value in Cold-Spots with respect to (d) Earthquake and Total Power Outage Events, (e) Hurricane and Total Power Outage Events, and (f) Tornado and Power Outage Events.

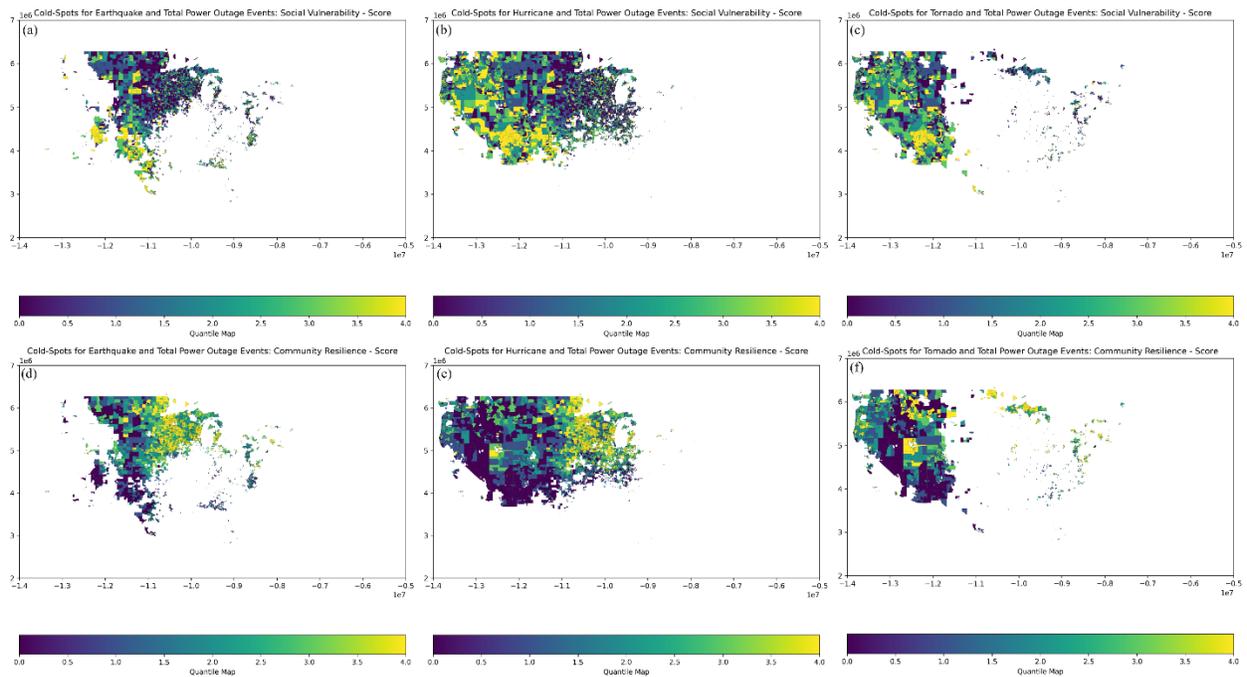

**Fig. 2A.** Social Vulnerability in Cold-Spots with respect to (a) Earthquake and Total Power Outage Events, (b) Hurricane and Total Power Outage Events, and (c) Tornado and Power Outage Events. Community Resilience in Cold-Spots with respect to (d) Earthquake and Total Power Outage Events, (e) Hurricane and Total Power Outage Events, and (f) Tornado and Power Outage Events.